\documentclass[aps,prl,superscriptaddress,amsmath,amsfonts,amssymb,showpacs,reprint,longbibliography,floatfix]{revtex4-2}
\usepackage[utf8]{inputenc}
\usepackage[T1]{fontenc}
\usepackage{mathtools}
\usepackage{newtxtext,newtxmath}
\usepackage{amssymb}
\usepackage{amsmath}
\usepackage{microtype}
\usepackage{graphicx}
\usepackage[dvipsnames,table]{xcolor}
\usepackage{natbib}
\usepackage{hyperref}
\usepackage{physics}
\usepackage{bm}
\usepackage{bbm}
\usepackage[version=4]{mhchem}
\usepackage[percent]{overpic}
\usepackage[capitalise]{cleveref}
\usepackage{nicematrix}
\usepackage{longtable}
\usepackage{array}
\usepackage{dcolumn}
\usepackage{comment}
\usepackage{tikz}
\usepackage[compat=1.1.0]{tikz-feynman}
\tikzfeynmanset{warn luatex=false}

\definecolor{cblue}{RGB}{55,126,184}
\definecolor{ogreen}{RGB}{238,255,204}
\hypersetup{colorlinks=true,linkcolor=cblue,citecolor=cblue,urlcolor=cblue}

\newcommand{\transpose}{\intercal}
\newcommand{\bk}{\boldsymbol{k}}

\begin{document}

\title{Universal temperature-dependent power law excitation gaps in frustrated quantum spin systems harboring order-by-disorder}

\author{Alexander Hickey}
\affiliation{Department of Physics and Astronomy, University of Waterloo, Waterloo, Ontario, N2L 3G1, Canada}
\author{Jeffrey G. Rau}
\affiliation{Department of Physics, University of Windsor, 401 Sunset Avenue, Windsor, Ontario, N9B 3P4, Canada}
\author{Subhankar Khatua}
\affiliation{Department of Physics and Astronomy, University of Waterloo, Waterloo, Ontario, N2L 3G1, Canada}
\affiliation{Department of Physics, University of Windsor, 401 Sunset Avenue, Windsor, Ontario, N9B 3P4, Canada}
\affiliation{Institute for Theoretical Solid State Physics, IFW Dresden and W\"urzburg-Dresden Cluster of Excellence ct.qmat, Helmholtzstr. 20, 01069 Dresden, Germany}
\author{Michel J. P. Gingras}
\affiliation{Department of Physics and Astronomy, University of Waterloo, Waterloo, Ontario, N2L 3G1, Canada}

\begin{abstract}
When magnetic moments are subject to competing or frustrated interactions, continuous degeneracies that are not protected by any symmetry of the parent Hamiltonian can emerge at the classical (mean-field) level.
Such ``accidental'' degeneracies are often lifted by both thermal and quantum fluctuations via a mechanism known as order-by-disorder (ObD). 
The leading proposal to detect and characterize ObD in real materials, in a way that quantitatively distinguishes it from standard energetic selection, is to measure a small fluctuation-induced \emph{pseudo-Goldstone gap} in the excitation spectrum.
While the properties of this gap are known to leading order in the spin wave interactions, in both the zero-temperature and classical limits, the pseudo-Goldstone (PG)  gap in quantum magnets at finite temperature has yet to be characterized. 
Using non-linear spin wave theory, we compute the PG gap to leading order in a $1/S$ expansion at low temperature for a variety of frustrated quantum spin systems.
We also develop a formalism to calculate the PG gap in a way that \emph{solely} uses linear spin-wave theory, circumventing the need to carry out tedious quantum many-body calculations. 
We argue that, at leading order, the PG gap acquires a distinct power-law temperature dependence, proportional to either $T^{d+1}$ or $T^{d/2+1}$ depending on the gapless dispersion of the PG mode predicted at the mean-field level.
Finally, we examine the implications of these results for the pyrochlore oxide compound \ce{Er2Ti2O7}, for which there is compelling evidence of ObD giving rise to the experimentally observed long-range order.
\end{abstract}

\date{\today}

\maketitle

In systems of interacting degrees of freedom, both thermal and quantum fluctuations typically act to destabilize broken symmetries and long-range order. 
In certain condensed matter systems, these fluctuations can be so pronounced that they completely suppress order even at absolute zero temperature.
A striking example is liquid helium, which avoids freezing into a crystalline solid at atmospheric pressure due to strong quantum zero-point fluctuations~\cite{London1938}. 
Within the field of magnetism, the quest for quantum spin liquids---states that lack long-range magnetic order down to the lowest temperatures---has long been a central endeavor~\cite{Balents2010,gingras2014,Imai2016,Savary2017,Knolle2019}. One preeminent setting for this search considers highly frustrated magnetic systems where competing spin-spin interactions produce an exponentially large number of classically degenerate ground states~\cite{Villain_ZPhysB,moessner1998}.
This degeneracy leads to an extensive entropy that can prevent magnetic ordering down to absolute zero temperature, even in three-dimensional systems, giving rise to a classical spin liquid~\cite{moessner1998,lozano-gomez2023,lozano_atlas}. 
Such a classical perspective has motivated an intense theoretical and experimental exploration of quantum analogs of such systems.

An intriguing intermediate situation between energetically stabilized long-range magnetic order and fluctuation-driven spin liquid states happens in certain models of frustrated spin systems that possess a sub-exponentially large (in the system size) manifold of classical ground states. 
In these systems, the \emph{accidental} degeneracies are not protected by global symmetries of the spin Hamiltonian but rather emerge from the specific form of the spin-spin interactions \footnotetext[1]{Note that the existence of an accidental degeneracy by no means implies this intermediate scenario.}~\cite{Note1}.
Although no unique classical ground state (modulo global symmetries) is energetically favored at the mean-field level, thermal or quantum fluctuations can lift the degeneracy, stabilizing a subset of configurations within the manifold of classical ground states and cause long-range magnetic order---this is the celebrated phenomenon of \emph{order-by-disorder} 
(ObD)~\cite{villain1980,Shender1982,Henley1989}, a conceptual cornerstone of frustrated magnetism.

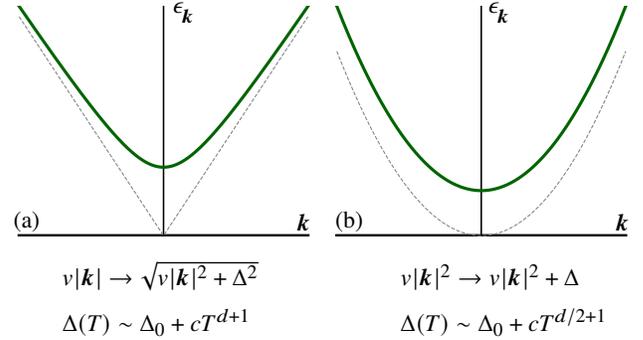
\begin{figure}[t]
    \centering
    \begin{overpic}[width=\columnwidth,trim=0 -3cm 0 0, clip]{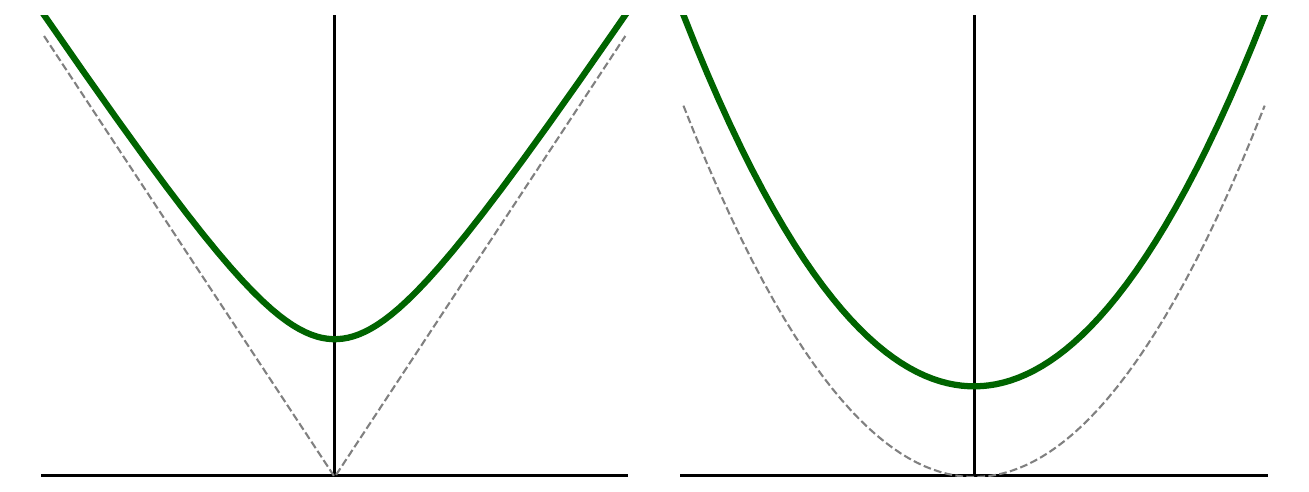}
    \put(10,7){$v|\vb*{k}| \to \sqrt{v|\vb*{k}|^2 + \Delta^2}$}
    \put(10,0){$\Delta(T) \sim \Delta_0 + cT^{d+1}$}
    \put(62,7){$v|\vb*{k}|^2 \to v|\vb*{k}|^2 + \Delta$}
    \put(62,0){$\Delta(T) \sim \Delta_0 + cT^{d/2+1}$}
    \put(2.5,16){(a)}
    \put(52,16){(b)}
    \put(46,15.5){$\vb*{k}$}
    \put(95,15.5){$\vb*{k}$}
    \put(27,49){$\epsilon_{\vb*{k}}^{}$}
    \put(75.5,49){$\epsilon_{\vb*{k}}^{}$}
    \end{overpic}
    \caption{Schematic picture depicting the spectrum of a PG modes for (a) type I with a linear dispersing mode, $\epsilon =v|{\bm k}|$, and (b) type II.
    with a quadratic dispersing mode, $\epsilon =v|{\bm k}|^2$. In both cases, the mode acquires a gap $\Delta$ 
    via the ObD mechanism, modifying the dispersion.
    Quantum fluctuations at zero temperature ($T=0$) generates a gap $\Delta_0$ (quantum ObD), which then
    \emph{increases} with $T$ as a power-law, $T^{d+1}$ and 
    $T^{d/2+1}$, for type I and type II mode, respectively. 
        }
    \label{fig:PG_gap}
\end{figure}

While ObD driven by thermal or quantum fluctuations has been well-established in an abundant number of theoretical models of frustrated spin
systems~\cite{villain1980,Henley1989,Henley1987,prakash1990,reimers1993,noculak2023,HickeyPRB,bergman2007,chern2010,Gvozdikova2011,mcclarty2014,Francini2024nematicR2,tessman1954,Shender1982,belorizky1980,chandra1988,kubo1991,sachdev1991,Chubukov_1991,chubukov1992,henley1994,Lecheminant1995,sobral1997,champion2003,baskaran2008,bernier2008,mulder2010,savary2012,zhitomirsky2012,Chernyshev2014,Rousochatzakis2015,rau2018,Placke2020,Chen2020,Liu2020,Khatua2021,khatua2024}, 
a long-standing question is whether ObD is realized in real magnetic materials and, if so, what experimental signatures can unambiguously demonstrate its role in stabilizing the observed long-range order. 
The conventional approach to experimentally confirming ObD has been rather indirect, proceeding by constructing a theoretical model of the material, demonstrating ObD within that framework, and then experimentally verifying a maximum number of the model’s predictions against the material properties~\cite{savary2012,Ross2014,sarkis2020,Elliot2021,kim1999,Brueckel1988}.
This program for identifying ObD in real materials raises two important concerns.
Firstly, it presents significant methodological challenges, as establishing detailed quantitative agreement between theory and experiment where ObD is operating necessitates intricate high-order quantum many-body perturbation theory calculations.
Secondly, even in cases where it can be successfully executed, the above program leaves one wanting for a deeper understanding of ObD, such as asking whether there may exist more direct experimental evidence for ObD in a material that do not rely on a back and forth ``dialogue’’ between theoretical calculations and the fitting of experimental data.

Two recent developments~\cite{rau2018,@khatua2023} have begun addressing the above two issues by considering the formation of an ObD-induced energy gap $\Delta$ in the otherwise gapless pseudo-Goldstone (PG) mode (i.e. spin-wave excitations) predicted by mean-field theory (see Fig.~\ref{fig:PG_gap}).
The first development is the observation that, for a given Hamiltonian model for a quantum spin-$S$ system, the curvatures of classical plus quantum zero-point energies computed at $O(S)$ are sufficient to determine the PG zero-temperature gap $\Delta_0$ \emph{exactly} to $O(S^0)$, 
without the necessity of performing involved self-energy calculations due to magnon-magnon interactions~\cite{rau2018}. 
The second development is the observation that frustrated classical spin systems, displaying thermal but no quantum ObD, feature a temperature-dependent gap generated by magnon interactions 
scaling like $\Delta(T) \propto T^\mu$ with $\mu= \frac{1}{2}$ ($\mu=1$) for linearly dispersing (quadratically dispersing) PG modes~\cite{@khatua2023}.
These two results beg for two important follow-up questions: ``What is the expected temperature-dependence of $\Delta(T)$ for a \emph{quantum} treatment of models with linearly or quadratically dispersing PG modes?'' and ``Can one compute such temperature dependence through a generalization of the framework of Ref.~\cite{rau2018} involving curvatures of the free energy density using only linear spin-wave theory?'' 
In this paper, we answer the second question in the affirmative, and show that the above exponent $\mu$ become  $d+1$ (linearly dispersing PG mode, $\epsilon_{\bk} \propto |{\bm k}|)$ and $d/2+1$  (quadratically dispersing PG mode, 
$\cramped{\epsilon_{\bk} \propto |{\bm k}|^2)}$, respectively (see Fig.~\ref{fig:PG_gap}).
In short, we have \emph{(i)} identified a ``smoking-gun'' experimental evidence for ObD in a spin system in the form of two ``universal'' types of ObD-generated temperature dependent gaps $\Delta(T)$ to the PG mode, and (\emph{ii}) propose a straightforward methodology to compute $\Delta(T)$ that is unburdened from complex finite-temperature quantum-many body calculations.

\emph{Spin-wave theory---}To investigate the thermal properties of the fluctuation-induced gap, we use the formalism of spin-wave theory, extending the calculations of Ref.~\cite{rau2018} to finite temperature to incorporate thermal fluctuation effects.
Formally, spin operators are mapped to Holstein-Primakoff 
bosons~\cite{Holstein1940} and subsequently expanded in powers of $1/S$, where $S$ is the spin quantum number~\cite{supp_material}.
Substituting this expansion into the Hamiltonian and keeping only the bilinear terms results in the linear spin-wave Hamiltonian
\begin{equation} \label{eqn:LSW}
    H_2 = \sum_{\vb*{k}}\sum_{\alpha\alpha'} \left[A_{\vb*{k}}^{\alpha \alpha'}a_{\vb*{k},\alpha}^\dagger a_{\vb*{k},\alpha'}^{} 
    + \frac{1}{2}\left(B_{\vb*{k}}^{\alpha \alpha'}a_{\vb*{k},\alpha}^\dagger a_{-\vb*{k},\alpha'}^\dagger + \text{H.c.}\right) \right],
\end{equation}
where $a_{\vb*{k},\alpha}^{}$ is a Holstein-Primakoff boson with wavevector $\vb*{k}$ on sublattice $\alpha$ of the magnetic unit cell, and the coefficients $A_{\vb*{k}}^{\alpha \alpha'}$ and $B_{\vb*{k}}^{\alpha \alpha'}$ depend on both the magnetic exchange interactions and the specific classical ground state configuration(s) considered~\cite{supp_material}. The quasiparticle dispersion is then determined by the eigenvalues of the (non-Hermitian) Bogoliubov-de Gennes matrix~\cite{Blaizot1986}
\begin{equation}
    \vb*{\sigma}_3 \vb*{M}_{\vb*{k}} =
    \vb*{\sigma}_3 
    \begin{pmatrix}
    \vb*{A}_{\vb*{k}} & \vb*{B}_{\vb*{k}} \\
    \vb*{B}_{\vb*{k}}^\dagger & \vb*{A}_{-\vb*{k}}^\transpose
    \end{pmatrix},
\end{equation}
where $\vb*{\sigma}_3 = \text{diag}(\mathbb{I},-\mathbb{I})$ is a Pauli matrix acting on the particle-hole degree of freedom.

Beyond linear spin-wave theory, the quasiparticle energies correspond to poles of the (retarded) single-magnon propagator
\begin{equation} \label{eqn:GreensFuntion}
    \vb*{G}^{\rm R}(\vb*{k},\omega) = 
    \left[\omega + i0^+ -\vb*{\sigma}_3\left( S \vb*{M}_{\vb*{k}} + \vb*{\Sigma}^{\rm R}(\vb*{k},\omega)\right)\right]^{-1} 
    \vb*{\sigma}_3 ,
\end{equation}
where $ \vb*{\Sigma}^{\rm R}(\vb*{k},\omega)$ is the retarded (thermal) self-energy matrix. We note that the matrix in Eq.~\eqref{eqn:GreensFuntion} has both sublattice indices and indices that track the normal and anomalous 
components \footnotetext[2]{Here, ``normal'' refers to the number-conserving blocks of Eq.~\eqref{eqn:LSW} (i.e. terms proportional to $a^\dagger a$ and $aa^\dagger$) and ``anomalous'' refers to the number-non-conserving blocks (i.e. terms proportional to $a^\dagger a^\dagger$ and $aa$).}~\cite{Blaizot1986,rau2018,Note2}. 
We may then include leading order effects of spin-wave interactions perturbatively by computing the self-energy to $O(S^0)$ in the $1/S$ expansion~\cite{Rastelli2013}.
This contribution to the self-energy can be represented as the sum of three one-particle irreducible diagrams \cite{NegeleOrland}
\begin{equation}
\vb*{\Sigma}^{\rm R} = 
\begin{tikzpicture}[baseline = -1.25mm]
\newcommand \Lpanel {\columnwidth/4};
\newcommand \whitespace {\Lpanel/20};
\begin{feynman}[inline=(al1.base)]
    \vertex (al1) {\( \)};
    \vertex [dot,right=\Lpanel/3 of al1] (vl1) {};
    \vertex [dot,right=\Lpanel/3 of vl1] (vr1) {};
    \vertex [above right=1.414213\Lpanel/6 of vl1] (mt1);
    \vertex [below right=1.414213\Lpanel/6 of vl1] (ml1);
    \vertex [right=\Lpanel of al1] (ar1) {\( + \)};
    \diagram*{
        (al1) -- [plain] (vl1),
        (vl1) -- [plain, quarter left] (mt1)
              -- [plain, quarter left] (vr1)
              -- [plain, quarter left] (ml1)
              -- [plain, quarter left] (vl1),
        (vr1) -- [plain] (ar1)
    };
    \vertex [right=\whitespace of ar1] (al2) {\( \)};
    \vertex [right=\Lpanel of al2] (ar2) {\( + \)};
    \vertex [dot,right=\Lpanel/2 of al2] (vb2) {};
    \vertex [dot,above=\Lpanel/6 of vb2] (vt2) {};
    \vertex [above left=1.414213\Lpanel/6 of vt2]  (ll2);
    \vertex [above right=1.414213\Lpanel/6 of vt2] (rl2);
    \vertex [above left =1.414213\Lpanel/6 of rl2] (tl2);
    \diagram*{
        (al2) -- [plain] (vb2)
              -- [plain] (ar2) ,
        (vb2) -- [plain] (vt2)
              -- [plain, quarter right] (rl2)
              -- [plain, quarter right] (tl2)
              -- [plain, quarter right] (ll2)
              -- [plain, quarter right] (vt2)
    };
    \vertex [right=\whitespace of ar2] (al3) {\( \)};
    \vertex [right=\Lpanel of al3] (ar3) {\( \)};
    \vertex [dot,right=\Lpanel/2 of al3] (vb3) {};
    \vertex [above left= 1.414213\Lpanel/6 of vb3] (mcl3);
    \vertex [above right=1.414213\Lpanel/6 of vb3] (mcr3);
    \vertex [above left=1.414213\Lpanel/6 of mcr3] (mt3);
    \diagram*{
        (al3) -- [plain] (vb3)
        -- [plain] (ar3) ,
        (vb3) -- [plain, quarter right] (mcr3)
             -- [plain, quarter right] (mt3)
             -- [plain, quarter right] (mcl3)
             -- [plain, quarter right] (vb3),
    };
\end{feynman}
\end{tikzpicture},
\end{equation}
constructed from both three- and four-magnon interaction vertices.
In general, these diagrams carry a conserved wavevector and frequency $(\bk,\omega)$,  and sublattice indices, and include both normal and anomalous propagators~\cite{Note2,supp_material}.
Note that in order to incorporate thermal effects into the self-energy, we must take the limit $1/S \to 0$ while keeping $T/S$ fixed~\cite{supp_material}.

Next, we consider  a gapless PG mode, which falls into one of two categories depending on its spectral properties, referred to as ``type I'' and ``type II''~\cite{Watanabe2012,rau2018}.
Throughout this paper, we assume that this mode is located at the Brillouin zone center ($\vb*{k}=\vb*{0}$), although this assumption can be relaxed in general.
At the level of linear spin-wave theory, a type I PG mode has linear dispersion,  $\epsilon_{\bk} \propto |\vb*{k}|$,  while the type II mode disperses quadratically, $\epsilon_{\bk} \propto |\vb*{k}|^2$, as depicted in Fig.~\ref{fig:PG_gap}.
For a type I mode, there is a single linearly independent eigenvector $\vb*{V}_0$ corresponding to a zero eigenvalue. In this case, the PG gap appears at $O(S^{1/2})$ as
\begin{equation}\label{eqn:typeIgap}
    \Delta(T) = S^{1/2}\sqrt{
    \vb*{V}_0^\dagger
    \left[
\vb*{\Sigma}^{\rm R}_0(T)\vb*{\sigma}_3 \vb*{M}_{\vb*{0}} 
    +  \vb*{M}_{\vb*{0}} \vb*{\sigma}_3\vb*{\Sigma}^{\rm R}_0(T) 
    \right]
    \vb*{V}_0^{} } ,
\end{equation}
where $\vb*{\Sigma}^{\rm R}_0(T) \equiv \vb*{\Sigma}^{\rm R}(\vb*{0},0)$ is introduced to make the temperature dependence of the self-energy explicit.
For a type II mode, there are two linearly independent eigenvectors, $\vb*{V}_0$ and $\vb*{W}_0$, corresponding to a zero eigenvalue. 
In this case, the PG gap appears at $O(S^{0})$ as
\begin{equation}\label{eqn:typeIIgap}
    \Delta(T) = S^{0}\sqrt{
    \left(
    \vb*{V}_0^\dagger \vb*{\Sigma}^{\rm R}_0(T)\vb*{V}_0^{} 
    \right)^2 
    - \Big| 
    \vb*{V}_0^\dagger \vb*{\Sigma}^{\rm R}_0(T) \vb*{W}_0^{} 
    \Big|^2} .
\end{equation}
We refer the interested reader to the Supplementary Material of Ref.~\cite{rau2018} for a detailed derivation of Eqs.~(\ref{eqn:typeIgap},\ref{eqn:typeIIgap}). 
With the PG gap, the low energy dispersion now takes the form $\sqrt{v^2|\vb*{k}|^2 + \Delta^2}$ for a type I mode and $v|\vb*{k}|^2 + \Delta$ for a type II mode, as depicted in Fig.~\ref{fig:PG_gap}. 
The gap then acquires a temperature dependence $\Delta \rightarrow \Delta(T)$ from the self-energy $ \vb*{\Sigma}^{\rm R}(\vb*{k},\omega)$ in Eqs.~(\ref{eqn:typeIgap},\ref{eqn:typeIIgap}).

\emph{Curvature formula---}In previous work characterizing the PG gap in both the quantum zero-temperature~\cite{rau2018} and classical finite-temperature~\cite{@khatua2023} scenarios, it was argued that the gap at leading order can be calculated exactly by computing the curvature of the linear spin-wave free energy.
Here, we develop a similar formalism analogous to Refs.~\cite{rau2018,@khatua2023} that allows us to calculate the PG gap at $O(S^0)$ using only the linear spin-wave energy eigenvalues at $O(S)$, circumventing the need to carry out tedious calculations of the magnon self-energy.

The derivation of such a formula resembles the zero-temperature proof in Ref.~\cite{rau2018}, which can be generalized to finite temperature. 
Consider a local reference frame $(\vu*{e}_{x,\alpha}^{},\vu*{e}_{y,\alpha}^{},\vu*{e}_{0,\alpha}^{})$ where $\vu*{e}_{0,\alpha}^{}$ is the classical ordering direction, $\vu*{e}_{y,\alpha}$ is the soft mode direction, and $\vu*{e}_{x,\alpha}^{} = \vu*{e}_{y,\alpha}^{} \times \vu*{e}_{0,\alpha}^{}$.
We may parametrize the soft mode by an angle $\phi$ (along with its canonically conjugate pair $\theta$) and define the rotated Hamiltonian $\mathcal{H}(\phi,\theta) = U(\phi,\theta)^\dagger H U(\phi,\theta)$ where $U(\phi,\theta)$ is a unitary transformation that generates the soft modes of the spin configurations. 
Formally, this modifies the local frame as
\begin{equation}
    \begin{pmatrix}
    \vu*{e}_{x,\alpha}^{}(\phi,\theta) \\
    \vu*{e}_{y,\alpha}^{}(\phi,\theta) \\
    \vu*{e}_{0,\alpha}^{}(\phi,\theta)
    \end{pmatrix}
    = 
    \begin{pmatrix}
    \cos \phi & -\sin \phi \sin \theta & \sin \phi \cos \theta \\
    0 & \cos \theta & \sin \theta \\
    -\sin \phi  & -\cos \phi \sin \theta & \cos \phi \cos \theta
    \end{pmatrix}
    \begin{pmatrix}
    \vu*{e}_{x,\alpha}^{} \\
    \vu*{e}_{y,\alpha}^{} \\
    \vu*{e}_{0,\alpha}^{}
    \end{pmatrix}.
\end{equation}
We subsequently expand the rotated Hamiltonian to $O(S)$ to obtain
\begin{align}
    \mathcal{H}(\phi,\theta) &= S(S+1)N\epsilon_{\rm cl}(\theta) + SN\epsilon_{\rm qu}(\phi,\theta) \nonumber \\
    &+ S \sum_{\vb*{k},\alpha}\epsilon_{\vb*{k},\alpha}^{}(\phi,\theta) b_{\vb*{k},\alpha}^\dagger b_{\vb*{k},\alpha}^{} ,
    \label{eqn:RotatedHamiltonian}
\end{align}
where $S^2 \epsilon_{\rm cl}$ is the classical ground state energy (per spin) and $S(\epsilon_{\rm cl}+\epsilon_{\rm qu})$ is the quantum zero-point energy (per spin), and $S\epsilon_{\vb*{k},\alpha}^{}$ is the linear spin-wave dispersion~\cite{supp_material}. 
Equation~\eqref{eqn:RotatedHamiltonian} effectively parametrizes the magnons with respect to the zero-mode subspace.
For a type I mode, only $\phi$ corresponds to a soft mode at the classical level, so the classical configuration energy depends on $\theta$. 
When $\theta \neq 0$, the spin configuration is classically unstable, implying that the zero-point and spin-wave energies are not well-defined.
For a type II mode, both angles are classically soft, so $\epsilon_{\rm cl}(\theta) = \epsilon_{\rm cl}(0)$.
In general, the free energy per spin at $O(S)$ is
\begin{align}
    f(\phi,\theta) &= S(S+1)\epsilon_{\rm cl}(\theta) + S\epsilon_{\rm qu}(\phi,\theta) \nonumber \\
    &+ \frac{k_{\rm B}T}{N} \sum_{\vb*{k},\alpha} \ln \left(1-e^{-S\epsilon_{\vb*{k},\alpha}(\phi,\theta)/k_{\rm B}T}\right).
    \label{eqn:LSWFreeEnergy}
\end{align}

We may now establish an equivalence between the effective Hamiltonian to $O(S^1)$ in Eq.~\eqref{eqn:RotatedHamiltonian} and the PG gap using a sum rule for the magnon spectral function 
\begin{equation} \label{eqn:SumRule}
     \frac{1}{SN}\expval{\left(\pdv[2]{\mathcal{H}}{\lambda_\mu}{\lambda_\nu}\right)_0} = \vb*{U}_\mu^\dagger \vb*{\sigma}_3 \left[ \int_{-\infty}^{\infty} \dd{\omega} \omega \; \vb*{\mathcal{A}}(\vb*{0},\omega) \right] \vb*{\sigma}_3 \vb*{U}_\nu ,
\end{equation}
where $(\cdots)_0$ is used to denote evaluation at $\cramped{\phi=\theta = 0}$, $\cramped{\mu,\nu = \phi,\theta}$ label the two angles, $\lambda_\phi = \phi$, $\lambda_\theta = \theta$, and $\vb*{\mathcal{A}}(\vb*{0},\omega) \equiv \frac{1}{2i}\left(\vb*{G}^{\rm R}(\vb*{0},\omega)-\vb*{G}^{\rm R}(\vb*{0},\omega)^\dagger\right)$ is the single-magnon spectral function evaluated at temperature $T$. The vectors $\vb*{U}_\mu$ are defined as $\vb*{U}_\phi \equiv i(\vb*{W}_0-\vb*{V}_0)/\sqrt{2}$, $\vb*{U}_\theta \equiv (\vb*{W}_0+\vb*{V}_0)/\sqrt{2}$, where $\vb*{V}_0$ and $\vb*{W}_0$ are vectors that span the zero-mode subspace in linear spin wave theory~\cite{supp_material,rau2018}. 
Here, $\langle \cdots\rangle$ denotes the thermal expectation value at a fixed temperature $T$.

The right-hand side of Eq.~\eqref{eqn:SumRule} is related to the PG gap at $O(S^0)$, while the left-hand side is related to curvature of the linear-spin wave energies at $O(S)$ and $O(S^2)$~\cite{supp_material}, paving the way for bypassing the need of performing self-energy calculations. 
In particular, the thermal gap at leading order is given by the formula
\begin{equation} \label{eqn:GapCurvature}
    \Delta(T) = 
    \begin{cases}
    S^{1/2}\sqrt{\left(\pdv[2]{\epsilon_{\rm cl}}{\theta}\right)_0g_{\phi\phi}^{}} & (\text{type I}) \\
    S^0 \sqrt{g_{\phi\phi}^{}g_{\theta\theta}^{}-g_{\phi\theta}^2} & (\text{type II}), 
    \end{cases}
\end{equation}
where 
\begin{equation} \label{eqn:CurvatureOfH}
    g_{\mu\nu}(T) \equiv \frac{1}{S}\left[ \left(\pdv[2]{f}{\lambda_\mu}{\lambda_\nu}\right)_0 + K_{\mu\nu} \right],
\end{equation}
and
\begin{equation}\label{eqn:ExtraTerm}
    K_{\mu\nu} \equiv 
    \frac{S^2}{4k_{\rm B}TN}\sum_{\vb*{k},\alpha}
    \left( \pdv{\epsilon_{\vb*{k},\alpha}^{}}{\lambda_\mu}\right)_0 
    \left( \pdv{\epsilon_{\vb*{k},\alpha}^{}}{\lambda_\nu}\right)_0
    \csch^2\left(\frac{S\epsilon_{\vb*{k},\alpha}^{}}{2k_{\rm B}T}\right) .
\end{equation}
As $T \to 0$, $\frac{S^2}{k_{\rm B}T}\csch^2\left(\frac{S\epsilon_{\vb*{k},\alpha}^{}}{2k_{\rm B}T}\right) \to \frac{4k_{\rm B}T}{\epsilon_{\vb*{k},\alpha}^{2}}$ for small $\epsilon_{\vb*{k},\alpha}^{}$, while $\csch^2\left(\frac{S\epsilon_{\vb*{k},\alpha}^{}}{2k_{\rm B}T}\right) \to 0$ exponentially otherwise.
Therefore $K_{\mu\nu}$ vanishes in the zero temperature limit.
At high temperature, $K_{\mu \nu} \propto T$, so this term does not affect the scaling proposed in Ref.~\cite{khatua2023} for the classical PG gap.
Note that since $T/S$ is kept fixed in perturbation theory, $g_{\mu\nu}$ is formally of $O(S^0)$ whenever both $\mu$ and $\nu$ correspond to soft modes.
The derivatives in Eq.~\eqref{eqn:ExtraTerm} can be readily calculated using the Hellmann-Feynman theorem~\cite{Feynman1939}. 
More details regarding the proof of Eq.~\eqref{eqn:GapCurvature} and the calculation of Eq.~\eqref{eqn:ExtraTerm} are provided in the Supplemental 
Material~\cite{supp_material}.
The formula for the PG gap in Eq.~\eqref{eqn:GapCurvature} is a finite-temperature generalization of the curvature formula in Ref.~\cite{rau2018}.

\begin{figure}[t!]
    \centering
    \begin{overpic}[width=1.0\columnwidth]{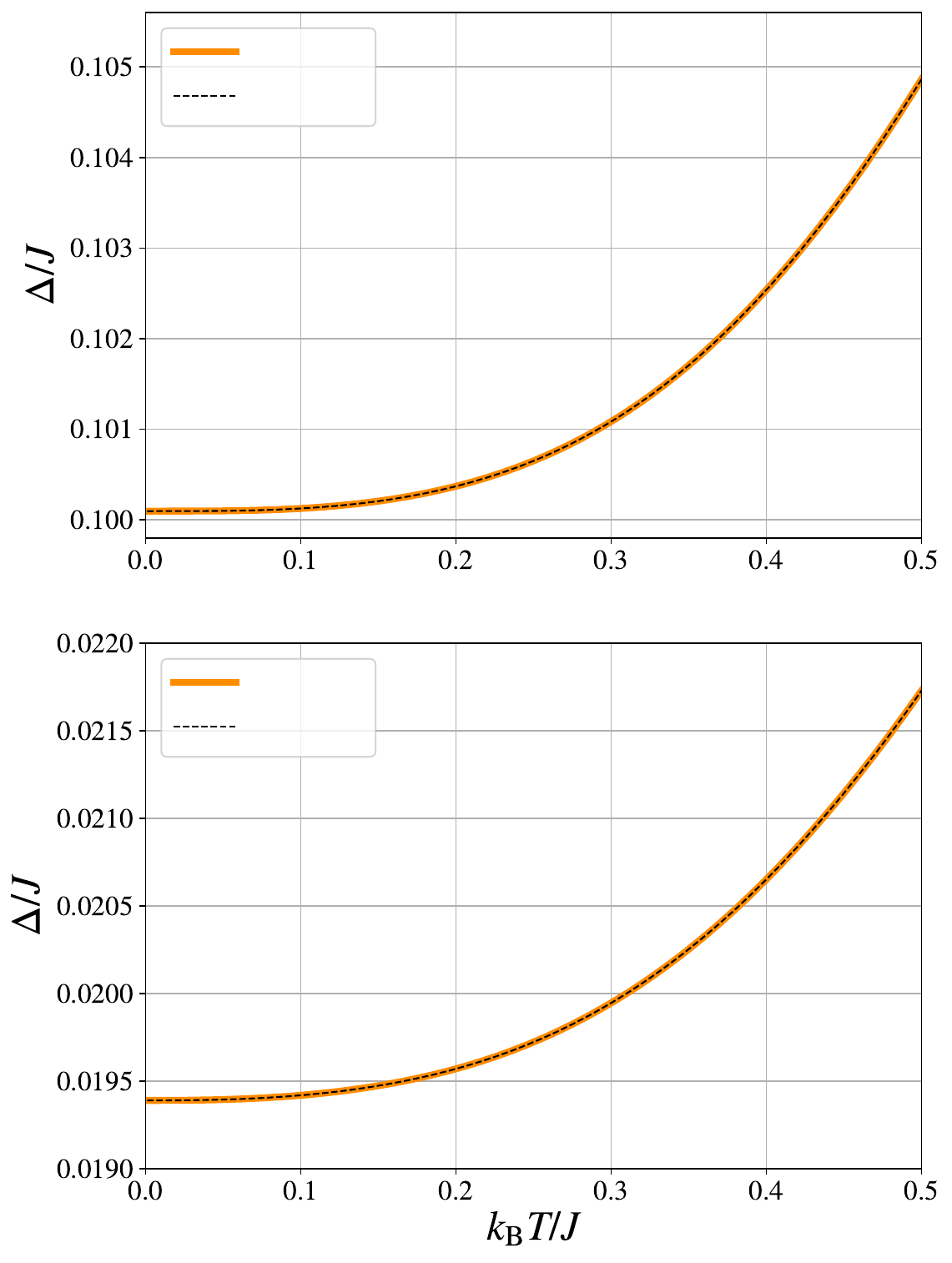}
    \put(20.25,95.00){Eq.~\eqref{eqn:typeIgap}}
    \put(20.25,91.50){Eq.~\eqref{eqn:GapCurvature}}
    \put(20.25,45.00){Eq.~\eqref{eqn:typeIIgap}}
    \put(20.25,41.50){Eq.~\eqref{eqn:GapCurvature}}
    \put(19,75){\large $\Delta(T)-\Delta_0 \propto T^{4}$}
    \put(19,29){\large $\Delta(T)-\Delta_0 \propto T^{5/2}$}
    \put(69.25,96.5){(a)}
    \put(69.25,46){(b)}
    \end{overpic}
    \caption{PG gap for the Heisenberg-compass model Eq.~\eqref{eqn:3DCompassModel} on a simple cubic lattice with (a) $\xi = 0$ (type I mode) and (b) $\xi = 1$ (type II mode). In both cases, we use $K/J = 0.5$ and $S=1/2$.}
    \label{fig:CompassGap}
\end{figure}
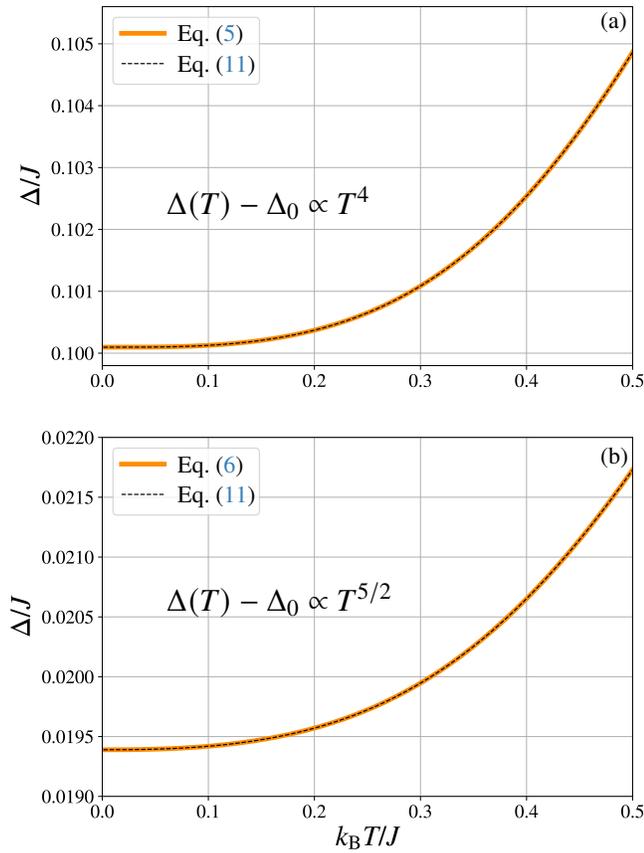

\emph{Models---} As a first example, we consider a ferromagnetic Heisenberg-compass model on the simple cubic lattice, defined by the Hamiltonian
\begin{align}
    H &= 
    -\sum_{\vb*{r}} \Biggl[
    J\sum_{\vb*{\delta} = \vb*{x},\vb*{y},\vb*{z}} 
    \vb*{S}_{\vb*{r}}^{} \cdot \vb*{S}_{\vb*{r}+\vb*{\delta}}^{} \nonumber \\
    &+ K\left( 
    S_{\vb*{r}}^x S_{\vb*{r}+\vb*{x}}^x 
    + S_{\vb*{r}}^y S_{\vb*{r}+\vb*{y}}^y 
    + \xi S_{\vb*{r}}^z S_{\vb*{r}+\vb*{z}}^z \right)
    \Biggr], \label{eqn:3DCompassModel}
\end{align}
where $\vb*{S}_{\vb*{r}}$ is a spin-$S$ operator at site $\vb*{r}$ and $\vb*{\delta}=\vb*{x},\vb*{y},\vb*{z}$ are the nearest-neighbor bonds, and $J,K>0$. For $K = 0$, Eq.~\eqref{eqn:3DCompassModel} is simply the Heisenberg ferromagnet, with collinear ground states related by a global $SU(2)$ symmetry. When $K>0$, there is no longer an exact continuous symmetry. 
In the following, we consider the cases of $\xi = 0$ and $\xi = 1$.

When $\xi=0$, the classical ground state corresponds to a collinear ferromagnet with magnetization along any direction in the $\vu*{x}-\vu*{y}$ plane. 
In contrast to the ($K=0$) Heisenberg model, this degeneracy is \emph{accidental}. 
This accidental $U(1)$ degeneracy is lifted by quantum and thermal fluctuations to select one of four magnetization directions along the $\pm \vu*{x},\pm \vu*{y}$ directions. 
In this case, the associated PG mode is of type I.

When $\xi=1$, the classical ground state is a collinear ferromagnet with arbitrary magnetization direction, signaling an accidental $O(3)$ degeneracy. 
The degeneracy is lifted by fluctuations to select one of six magnetization directions along the cubic axes $\pm \vu*{x},\pm \vu*{y},\pm \vu*{z}$. 
In this latter case, the associated PG mode is of type II \cite{supp_material}.

The PG gap at $O(S^0)$ for Eq.~\eqref{eqn:3DCompassModel} is depicted in Fig.~\ref{fig:CompassGap} over a range of  temperature in the case of both a type I ($\xi = 0$) and type II ($\xi = 1$) mode. 
We find an excellent quantitative agreement between the gap calculated using the curvature formula in Eq.~\eqref{eqn:GapCurvature} and the self-energy calculation of Eqs.~(\ref{eqn:typeIgap},\ref{eqn:typeIIgap}).
We emphasize that these two calculations are carried out independently of one another, and there are no free parameters introduced to make them agree. 
We note that the gap calculation is relatively simple for this model, as the lack of three-magnon interactions implies that $K_{\mu\nu}=0$ in Eq.~\eqref{eqn:CurvatureOfH}~\cite{supp_material}.
In both cases, there is a zero-temperature contribution due to the gap $\Delta_0$ induced by quantum ObD (i.e. zero-point fluctuations)~\cite{rau2018}. 
At low-temperature, the leading thermal contribution to the gap is proportional to $T^{4}$ for a type I mode and $T^{5/2}$ for a type II mode.

Next, we discuss an application to the XY pyrochlore antiferromagnet \ce{Er2Ti2O7}, which is arguably one of the best material candidates for ObD~\cite{savary2012,Ross2014,Petit2014}.
The spin-orbit entangled electronic states of \ce{Er^3+} are subject to a highly anisotropic crystal-field producing a ground doublet described using an effective (pseudo) spin-$\frac{1}{2}$ with  the Er$^{3+}$$-$E$^{3+}$ interactions described by an anisotropic exchange-like model~\cite{savary2012,Ross2011,Rau2019_review}
\begin{align}
    H &= \sum_{\langle i,j \rangle} \Bigl[
    J_{zz} S_i^z S_j^z 
    - J_{\pm}\left(S_i^+ S_j^- + S_i^- S_j^+ \right) \nonumber \\
    &+ J_{\pm \pm} \left( \gamma_{ij} S_i^+ S_j^+ + \text{H.c.} \right) \nonumber \\
    &+ J_{z \pm} \left(\zeta_{ij}\left[ S_i^z S_j^+ +  S_i^+ S_j^z\right] + \text{H.c.} \right)
    \Bigr] .
    \label{eq:Haniso}
\end{align}
Here $\langle i,j \rangle$ denotes the sum over nearest-neighbor bonds, and $\gamma_{ij},\zeta_{ij}$ are bond-dependent phase factors uniquely determined by the lattice geometry and symmetries of the single-ion wavefunctions~\cite{Rau2019_review}. 
The four nearest-neighbour $J_{uv}$ couplings in Eq.~\eqref{eq:Haniso} have been fitted to reproduce inelastic neutron scattering data on \ce{Er2Ti2O7}~\cite{savary2012}.
The classical ground states are non-collinear antiferromagnetic configurations of spins lying in the local XY planes perpendicular to the local $\langle 111 \rangle$ cubic axes of the pyrochlore lattice~\cite{Rau2019_review}, parametrized by an accidental $U(1)$ 
degeneracy~\cite{mcclarty2014,yan2017}.
Below $T_{\rm c} \approx 1.17$ K~\cite{champion2003}, ObD selects one of the six ``$\psi_2$'' states, leading to a type I PG mode~\cite{savary2012,Rau2019_review}.

\begin{figure}[t!]
    \centering
    \begin{overpic}[width=\columnwidth]{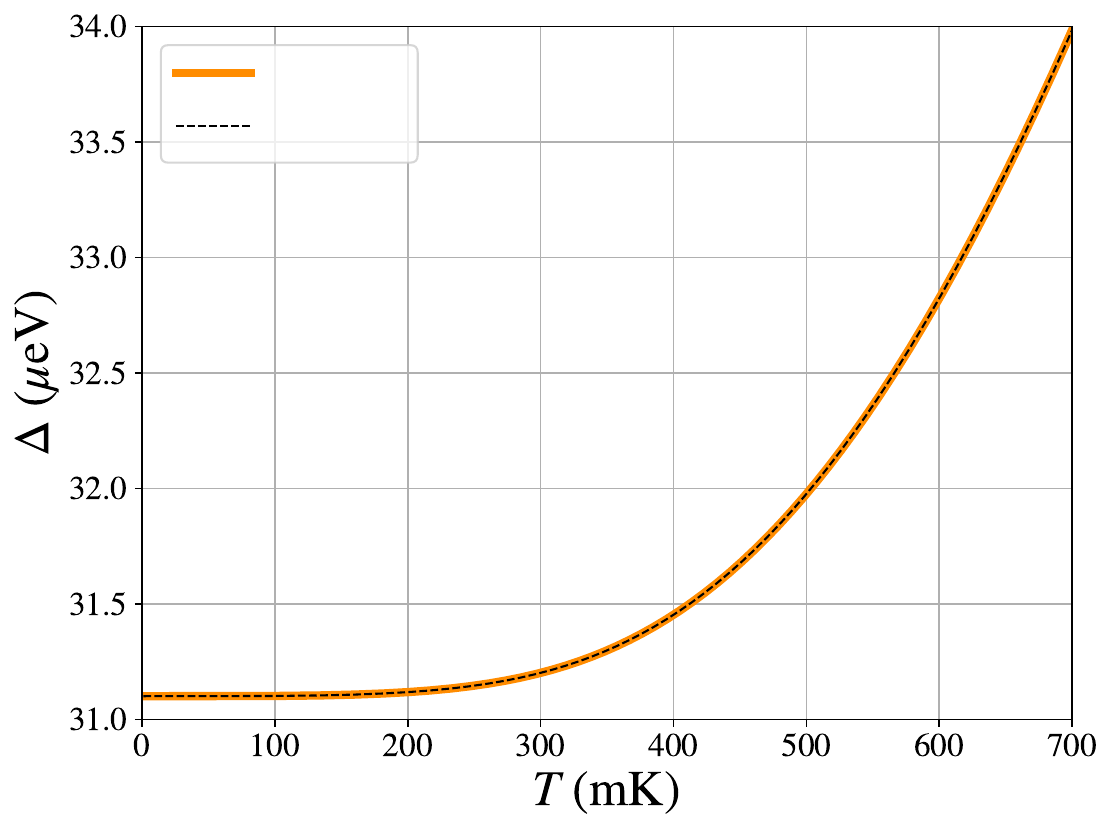}
        \put(24.5,67){Eq.~\eqref{eqn:typeIgap}}
        \put(24.5,62.5){Eq.~\eqref{eqn:GapCurvature}}
        \put(27,35){\large $\Delta(T)-\Delta_0 \propto T^{4}$}
    \end{overpic}
    \caption{PG gap for the pyrochlore antiferromagnet \ce{Er2Ti2O7}, calculated using the exchange couplings from Ref.~\cite{savary2012}. The classical ground states are parametrized by an accidental $U(1)$ degeneracy, corresponding to a type I mode. The gap is plotted up to $T=700 \text{ mK}$.}
    \label{fig:Er2Ti2O7Gap}
\end{figure}

The PG gap for \ce{Er2Ti2O7} is depicted in Fig.~\ref{fig:Er2Ti2O7Gap} over a range of temperatures up to $T=700 \text{mK}$. 
We find a zero-temperature contribution to the gap of $\Delta_0 = 31.1$ $\mu$eV, consistent with the gap calculated in Ref.~\cite{rau2018}, and somewhat smaller than the experimentally determined values of $43 \text{ }\mu\text{eV}$~\cite{Lhotel2017} 
and $53 \text{ }\mu\text{eV}$~\cite{Ross2014}. 
At low-temperature, the gap scales proportional to $T^4$, similar to the type I mode in Fig.~\ref{fig:CompassGap}(a).

\emph{Discussion---}The curvature formula in Eq.~\eqref{eqn:GapCurvature} directly implies that the PG gap satisfies a universal scaling relation as a function of temperature.
By noting that the spin-wave dispersion is linear ($\epsilon_{\vb*{k}} \sim |\vb*{k}|$) for a type I mode and quadratic ($\epsilon_{\vb*{k}} \sim |\vb*{k}|^2$) for a type II mode, a straightforward dimensional analysis of Eq.~\eqref{eqn:CurvatureOfH} implies that the leading thermal correction generically satisfies
\begin{equation} \label{eqn:UniversalScaling}
    \Delta(T) - \Delta_0 \propto 
    \begin{cases}
    S^{1/2}\left(\frac{T}{S}\right)^{d+1} & \text{(type I)} \\
    S^{0}\left(\frac{T}{S}\right)^{d/2+1} & \text{(type II)} ,
    \end{cases}
\end{equation}
where $d$ is the spatial dimension, and is consistent with the results in Figs.~\ref{fig:CompassGap} and \ref{fig:Er2Ti2O7Gap}.
We propose that this universal temperature-dependence of the gap \emph{increasing} with temperature at $0<T\ll T_{\textrm{c}}$, where $T_{\textrm{c}}$ is the critical temperature, can serve as experimental evidence for ObD in a spin systems that does not rely on fitting exchange couplings to a microscopic model.

The ability to calculate the PG gap using only linear spin-wave theory [Eq.~\eqref{eqn:GapCurvature}] provides a substantial simplification when compared to the tedious calculation of of the magnon self-energy, particularly in cases with non-collinear order.
In addition, the presence of a gapless mode in the linear spin-wave spectrum renders a direct calculation of the self-energy infrared divergent when $d<3$~\cite{Coleman1973}, drawing an analogy to the Hohenberg–Mermin–Wagner theorem~\cite{Mermin1966,Hohenberg1967,Berezinskii1970,Berezinskii1971}.
To circumvent this divergence, one could employ a more involved, self-consistent approach to calculate the self-energy, such as those described in Refs.~\cite{Loly1971,Fuhrman2012,Schick2022,@khatua2023,Gallegos2024}.
Alternatively, the curvature formula for the gap in Eq.~\eqref{eqn:GapCurvature} is well-defined for $d<3$, avoiding the need to regularize an infrared divergence in the first place.

Our observation that the PG gap obeys a universal, temperature-dependent scaling relation can serve as a framework to diagnose ObD in real materials in a way that is independent of the microscopic spin Hamiltonian describing the material.
Leading candidates where ObD may be present include the compounds \ce{Er2Ti2O7}~\cite{Ross2014,Petit2014,Lhotel2017}, \ce{CoTiO3}~\cite{Elliot2021}, \ce{Sr2Cu3O4Cl2}~\cite{kim1999}, and \ce{Fe2Ca3(GeO4)3}~\cite{Brueckel1988}. 
It remains unclear, however, whether the small gap measured in these materials is truly a PG gap, as it is possible that such a gap can arise from further neighbor exchange, anisotropic spin interactions or higher-order spin interactions such as a biquadratic interaction.
The experimental observation of a temperature-dependent gap such as in  Eq.~\eqref{eqn:UniversalScaling} would aid in resolving these questions.
The experimental details of how to resolve and characterize small thermal corrections to the PG gap remain an open problem.
For the pyrochlore material \ce{Er2Ti2O7}, the thermal correction to the gap up to $T=700 \text{ mK}$ is predicted to be $\Delta (700 \text{ mK}) - \Delta_0 = 2.9 \text{ }\mu$eV (see Fig.~\ref{fig:Er2Ti2O7Gap}).
This will likely be challenging to measuring experimentally with high resolution in temperature as this energy scale is at the lower end of what is currently accessible using high-resolution inelastic neutron backscattering measurements~\cite{neutron_back}.
In addition, there will be a crossover temperature, $T^*$, where thermal fluctuations and the progressive restoration of symmetry as $T_c$ is approached from below, will take over and make $\Delta(T)$ begin decreasing. 
The treatment of corrections to $\Delta(T)$ arising from pre-critical fluctuations is beyond the scope of this work, as we focus only on $\Delta(T)$ in the limit $T \ll T_{\rm c}$.
Nonetheless, our work provides a robust theoretical framework to guide future experimental efforts in this exciting field of research.

\emph{Note---}In the process of finalizing this manuscript, we became aware of a very recent preprint~\cite{Lin2025} that reports calculations of the PG gap at nonzero temperature similar to those presented here.

\emph{Acknowledgements---}We thank Michael Burke, Anton Burkov and Griffin Howson for useful technical discussions, and Jason Gardner, Bruce Gaulin and Romain Sibille for discussions 
about inelastic neutron backscattering. 
A. H.  acknowledges support from the NSERC of Canada CGS-D Scholarship. 
S. K. acknowledges financial support from the Deutsche Forschungsgemeinschaft (DFG, German Research Foundation) under Germany’s Excellence Strategy through the W\"urzburg-Dresden Cluster of Excellence on Complexity and Topology in Quantum Matter -- ct.qmat (EXC 2147, project-id 390858490).
The work at the U. of Waterloo and the U. of Windsor was supported by the NSERC of Canada (M. J. P. G. and J. G. R.) and the Canada Research Chair program (M. J. P. G., Tier 1).
This research was enabled in part by computing resources provided by the Digital Research Alliance of Canada.

\bibliography{refs}

\clearpage

\addtolength{\oddsidemargin}{-0.75in}
\addtolength{\evensidemargin}{-0.75in}
\addtolength{\topmargin}{-0.725in}

\newcommand{\addpage}[1]{
\begin{figure*}
  \includegraphics[width=8.5in,page=#1]{supp_mat.pdf}
\end{figure*}
}
\addpage{1}
\addpage{2}
\addpage{3}
\addpage{4}
\addpage{5}
\addpage{6}
\addpage{7}
\addpage{8}
\addpage{9}
\addpage{10}
\addpage{11}
\addpage{12}
\addpage{13}
\addpage{14}
\addpage{15}
\addpage{16}

\end{document}